# Shear-Banding Driven by Dynamic Associations in Soft Particle Glasses

Maddalena Mattiello[1], Paolo Edera[1], Jean-Marc Suau[2], Michel Cloitre[1]

[1] Molecular, Macromolecular Chemistry, and Materials, ESPCI Paris, PSL Research University, CNRS, 75005 Paris, France

[2] Arkema, ZI Lyon Nord, 35 Rue Ampère, 69730 Genay, France

Corresponding author: michel.cloitre@espci.psl.eu

## ABSTRACT

We develop associative colloids to investigate the role of dynamic associations on the physics of jammed suspensions. The particles are soft polymeric microgels functionalized with n-alkyl chains responsible for short-ranged hydrophobic associations when particles come into contact. Above the jamming transition, dynamic interactions are responsible for thermally-activated cage relaxation that controls the linear viscoelasticity, the yielding properties and the flow behavior. Jammed suspensions exhibit high elasticity and double yielding above a characteristic frequency $\omega_0$ whereas they behave like repulsive Soft Particle Glasses (SPG) below $\omega_0$. This frequency-dependent microscopic dynamic dictates the steady flow behavior. A constant stress plateau associated with shear-banding separates a low shear-rate region where associations increases the stress and a high shear-rate region where associations are fully disrupted and the stress coincides with that in repulsive SPGs. A microscopic model provides a physical interpretation of the shear-banding instability and a universal description of the flow properties in terms of the relaxation time and high frequency modulus. Our results demonstrate that the presence of dynamic attractive interactions strongly alters the viscoelasticity and the flow properties of jammed suspensions providing new routes to tune the rheology of soft jammed suspensions.

## I. INTRODUCTION

Over the past decade, considerable attention has been paid to soft particle glasses (SPGs) which consist of deformable particles at volume fractions beyond close-packing, where they are in contact and interact through purely repulsive elastic forces [1]. SPGs include many materials of practical and fundamental interest such as concentrated emulsions, microgel pastes, micelles, or star polymer solutions. Thanks to intense experimental, numerical, and theoretical studies a generic understanding of the structure and rheology of SPGs is now emerging. The concept of jamming borrowed from the statistical physics of athermal systems has proven relevant to describe and predict static properties such as the topology and elasticity of SPGs [2-5]. The rheological properties are characterized by the existence of a yield stress, $\sigma_y$, below which the particles are constrained in cages by elastic forces at rest, and above which they start moving over large distances [6]. After yielding, the flow curves that relate the shear stress $\sigma$ to the shear rate $\dot{\gamma}$ are generally well described by the Herschel-Bulkley (HB) equation $\sigma = \sigma_y + k\dot{\gamma}^n$ where $\sigma_y$, $k$, and $n$ are the yield stress, the consistency, and the shear-thinning exponent of order ½, respectively. This equation, which has been considered as an empirical yet useful form for a long time, has recently been justified on the basis of microscopic theories [7-9]. The manner that SPGs deform mechanically at the microscopic level has been a subject of debate but it is now established that SPGs with purely repulsive interactions flow homogeneously at steady state in the absence of wall slip [10].

Notwithstanding the fundamental and practical importance of SPGs where interactions are purely repulsive, many soft jammed materials encountered in industry involve attractive interactions. Depletion phenomena have been widely used to manipulate the behavior of complex mixtures containing hard sphere glasses [11], emulsions [12-14], concentrated star polymer solutions [15,16], and microgels [17,18], through the addition of depletants, which can be polymers, surfactants, or nanoparticles. When the colloids are permeable like microgels or star polymers, the depletants not only induce interparticle attractions but also provoke the deswelling of the particles by osmotic forces. In general, the structure and the properties of composite mixtures based on soft permeable colloids result from a subtle interplay between attractive interactions and osmotic shrinking [19]. A strategy to avoid deswelling is to functionalize the particles with complementary moieties that stick together when they are in contact. Following this idea, biomimetic adhesive emulsions have been successfully obtained by coating oil-in-water emulsion droplets with adhesive proteins [20,21], streptavidin biotin [22], or complementary DNA strands [23]. Functional microgels have also been prepared using host-guest and metal-ligand supramolecular chemistry [24]. Telechelic star polymer comprising asymmetric block copolymer arms with solvophilic inner and solvophobic outer blocks have been synthesized and investigated [25], where the strength of the interactions between the associating blocks is then varied by tuning the solvent quality through temperature changes. Despite these remarkable achievements, little is known on the impact of the short-range attractive interactions on the viscoelasticity and flow properties of the suspensions.

Theoretically, numerical simulations have interrogated the existence and the properties of the jamming transition in two-dimensional athermal suspensions of sticky disks. They predict that small amount of attraction fundamentally changes the nature of the transitions and the elastic properties with respect to the repulsive case [26-30]. Fragile solids similar to particulate gels form at a volume fraction $\phi_g$ that is much lower than the repulsive jamming transition that occurs at a volume fraction denoted $\phi_J$ hereafter. $\phi_g$ decreases when the strength of the interaction increases [29]. The jamming transition at $\phi_J$ corresponds to a rigidity percolation transition, which marks the arrest of the dynamics at all length scales, whereas $\phi_g$ corresponds to a connectivity transition [26]. In the gel-like regime $\phi_g < \phi < \phi_J$, the flow curves relating the shear stress to the applied shear rate have a non-monotonic shape leading to a mechanical instability that takes the form of localized shear bands [31,32]. In the jamming regime $\phi > \phi_J$, the velocity profiles present some weak inhomogeneities but the flow curves remain monotonic [32]. They are well described by standard Herschel-Bulkley equations and they can be rescaled onto a generic master curve using a single characteristic time that depends on the strength of the attraction [3030]. Recent 3D simulations of jammed SPGs interacting through a generalized Hertzian potential, short-range attractive forces, and elastohydrodynamic lubrication, have revealed local velocity fluctuations without any sign of shear-banding [33]. Shear-banding has been reported experimentally in various soft materials such as attractive emulsions [34-36], star glasses [37] and attractive microgels [38] but the connection at the microscopic level between experiments and simulations raises important questions.

The key difference between these models and experiments is that they consider that the dynamics of interparticle bonds only depend on hydrodynamic forces whereas many real systems involve dynamic interactions that have a finite lifetime beyond which they break and reform elsewhere. Such attractive interactions can lead to inhomogeneous flow in the jamming regime [39]. In this paper, we address in detail the connection between dynamic interactions and SPG rheology using two types of soft microgels synthesized with well-controlled emulsion polymerization techniques, which are similar in size and microstructure but differ in the interactions. The microgels of the first type have purely repulsive interactions; those of the second type are modified by alkyl chains, called stickers in the following, which are responsible for short-ranged hydrophobic associations. We find that attractive interactions do not affect the jamming concentration but strongly modify the material rheology. They have high elasticity and yield at large deformation at short time or high frequencies, but they respond like ordinary repulsive Soft Particle Glasses (SPG) at long times or low frequencies. The short and long time regimes are separated by a well-defined characteristic time that is associated to sticker dynamics. Associative SPGs flow homogeneously at low and high shear rates but not at intermediate shear rates where an interplay between wall slip and shear-banding is observed and rationalized. We demonstrate that the nonlinear flow properties are intimately linked to the viscoelastic properties in the quiescent state. In particular, the flow heterogeneities come from the strong thinning effect caused by the decrease of the lifetime of associations when the stickers are pulled out by the shearing flow.

## II. MATERIALS AND METHODS

### A. Synthesis of hydrophobically-modified microgels

Hydrophobically-modified microgels are synthesized by emulsion polymerization from ethyl acrylate ($59.8\times10^{-2}$ mol.mol$^{-1}$) methacrylic acid ($39.86\times10^{-2}$ mol.mol$^{-1}$), the crosslinker trimethylolpropane triacrylate ($2.0\times10^{-3}$ mol.mol$^{-1}$) and the macromonomer alkyl-*b*-polyoxyethylene methacrylate ($1.3\times10^{-3}$ mol.mol$^{-1}$). Macromonomers comprise a hydrophilic spacer made of 25 oxyethylene units (-$CH_2$-$CH_2$-O-) and an alkyl chain with 22 methylene units (-$C_{22}H_{45}$-) terminated by a methyl group. For reference, simple microgels with identical composition but without hydrophobic macromonomers are synthesized in the same conditions. In the following, hydrophobically-modified and simple microgels are called associative and repulsive microgels, respectively. In the jamming regime, associative microgels form associative Soft Particle Glasses (Fig. 1). They are synthesized at low pH using a semi-batch technique under starved conditions and continuous feeding of initiators and reactants [4]. The temperature and the feeding rate are optimized to yield a constant conversion rate and linear particle growth with time. These conditions are known to reduce heterogeneities in composition and crosslink density.

### B. Rheological measurements

We use a stress controlled rheometer (Anton Paar MCR 502) equipped with a cone and plate geometry (diameter: 25 mm; angle: 2°) and a single-gap Couette geometry (ref DG27) to explore the high shear rate domain. The shearing surfaces are sandblasted to prevent slip. After loading, the measuring system containing the suspensions are covered by a solvent trap to minimize water evaporation. The bottom plate is connected to a Peltier system that controls the temperature with an accuracy of 0.1 °C.

The linear viscoelastic properties of the suspensions are probed using small amplitude oscillatory shear (SAOS) experiments. The storage modulus $G'(\omega)$ and the loss modulus $G''(\omega)$ are determined by applying an oscillatory strain of small amplitude $\gamma_0 = 0.01$ and angular frequency $\omega$ varying between $10^{-3}$ and $10^{2}$ rad.s$^{-1}$. Large amplitude oscillatory shear experiments (LAOS), where periodic strain at fixed angular frequency and increasing amplitude is applied, are also performed to investigate yielding mechanisms at large deformation.

The flow properties are investigated using controlled-stress and controlled-rate experiments. In stress mode, the suspensions are subjected to a constant shear stress and the evolution of the strain and shear rate with time is recorded until steady state is reached. Conversely, in controlled rate mode, we apply a constant shear rate varying between $10^{3}$ s$^{-1}$ and $5\times10^{-4}$ s$^{-1}$ and we record the variations of the stress with time.

## C. Analysis of rheological data

### *1. Carreau-Yasuda model*

The variations of the viscosity versus the shear rate are modeled using a version of the Carreau-Yasuda model that includes the viscosity at low and infinite shear-rates, $\eta_\infty$ and $\eta_0$ respectively, and a single relaxation process [40]:

$$\eta(\dot{\gamma}) = \eta_\infty + \frac{\eta_0 - \eta_\infty}{\left[1+\left(\dot{\gamma}\tau_v\right)^a\right]^b} \tag{1}$$

$\tau_v$ is the longest relaxation time which marks the onset of shear thinning; *a* and *b* are two fitting parameters that characterize shear-thinning and the breadth of the transition. The Carreau–Yasuda model is implemented by minimizing the residuals between Eq. (1) and the experimental viscosity data using ordinary least-squares fitting.

### *2. Havriliak-Negami equation*

The variations of the viscoelastic moduli with the frequency are modeled using the Havriliak-Negami equation [41]:

$$\frac{G^*(\omega)-G_+}{G_+-G_-}=\left[\frac{1}{1+\left(i\omega\tau_0\right)^\alpha}\right]^\beta \tag{2}$$

*G*+ and *G*- are the high and low frequency moduli, respectively; $\tau_0 = 1/\omega_0$ is the relaxation time associated to the relaxation at $\omega_0$ that separates the high and low frequency regimes. $\alpha$ and $\beta$ are two parameters that characterize the breadth and the asymmetry of the relaxation at $\omega_0$, respectively. The general expression (2) can be developed to get:

$$G'(\omega)=G_+ +\left(G_- -G_+\right)\left[1+(\omega\tau_0)^{2\alpha}+2(\omega\tau_0)^\alpha\cos\left(\alpha\pi/2\right)\right]^{-1/2}\cos\theta \tag{3}$$

$$G''(\omega)=\left(G_- -G_+\right)\left[1+(\omega\tau_0)^{2\alpha}+2(\omega\tau_0)^\alpha\cos\left(\alpha\pi/2\right)\right]^{-1/2}\sin\theta \tag{4}$$

$$\theta=-\arctan\frac{(\omega\tau)^\alpha\sin\left(\alpha\pi/2\right)}{1+(\omega\tau)^\alpha\cos\left(\alpha\pi/2\right)} \tag{5}$$

The experimental data are analyzed as follows. To reduce the number of fitting parameters, we first determine the value of $\omega_0$ by fitting the maximum of the loss moduli to a second order polynomial. Moreover, the experimental data exhibit remarkable symmetry properties: $G'(\omega)$ has a center of symmetry at $(\omega_0, G'(\omega_0))$ and $G''(\omega)$ has a mirror symmetry with respect to $\omega=\omega_0$; this allows us to set $\beta=1$. We determine $\alpha$, *G*+ and *G*- by fitting the storage modulus data to the analytical forms (3) and (5). Finally $G''(\omega)$ is computed from (4) using these values of $\alpha$, *G*+ and *G*-.

**D. Fluorescence microscopy and particle-image velocimetry (micro-PIV)**

We use a custom-made micro-PIV equipment implemented on a commercial rheometer (Anton Paar MCR302) to measure the flow profiles inside the gap and access the true shear-rate ( $\dot{\gamma}$ ) that may be different from the apparent shear rate ( $\dot{\gamma}_a$ ) applied by the rheometer. The bottom plate is a quartz plate which has been roughened on the totality of its surface except on a small circular aperture through which the suspension can be observed from below. We use a fluorescence microscope consisting of a Mitutoyo unit (VMU V 378-505) equipped with a long working distance objective (M Plan Apo NIR 50x 378-825), a CCD camera (AVT Pike F-032B) and a fluorescence cube composed of a dichroic mirror (Leica T512-l), an excitation filter (Leica ET500-20x), and an emission filter (Leica ET535-30m). The microscope is mounted on a translation stage moved by a motor (Thorlabs APT stepper motor TST001). When the microscope is displaced vertically, the focal plane scans the entire gap of the cone and plate geometry. The suspension is seeded with spherical polystyrene microbeads labeled with Rhodamine B (Fluoro Max Red from Thermofischer; diameter = 0.5 μm). The samples are illuminated from below using a solid-state laser (λ = 532 nm; P = 800 mW) which is connected to the microscope through an optical fiber. Sequences of images are recorded using the software Hiris Automate commercially available from RD Vision and analyzed using an in-house software based on the cross-correlation technique. This equipment gives access to a range of velocities ranging from 1 $\mu m.s^{-1}$ to 20 $mm.s^{-1}$ with an accuracy better than 5 %; the spatial resolution is better than 5 μm.

## III. RESULTS AND DISCUSSION

### A. Microgel characterization

At low pH, repulsive and associative microgels are hard sphere-like particles with hydrodynamic radii measured using Dynamic Light Scattering (DLS) equal to 48±2 nm. The addition of sodium hydroxide causes the ionization of the carboxylic functions borne by the polymeric network and provokes the swelling of the microgels [42]. After swelling, we obtain optically clear suspensions at all concentrations without any sign of phase separation. Figure 1(A-B) shows schematics of repulsive and associative microgels. The hydrodynamic radii measured using DLS at high dilution are comparable: $R_h^{rep} = 335 \pm 12$ nm and $R_h^{ass} = 355 \pm 25$ nm, respectively. The particle radius distributions are weakly polydisperse with a polydispersity index of about 0.1. The form factors measured using Small Angle X-ray Scattering are superimposed over the whole range of wavevectors as shown in Fig. 1(C). Thus repulsive and associative microgels have the same microstructure, suggesting that macromonomers must be located at the surface of particles. Intraparticle associations would make associative microgels more compact, causing a decrease of the gyration and hydrodynamic radii. In conclusion hydrophobically-modified microgels can be viewed as soft polymeric networks decorated by short associative chains. The PEO spacers of the associative chains are soluble in water whereas the alkyl chain ends, which will be called stickers in the following, are in bad solvent and tend to associate when they are in contact. From the composition and radius of hydrophobically-modified microgels, we estimate that each particle comprises about 3700 associative chains on average. Assuming that they are statistically distributed on the surface, we estimate that they are separated by a distance of 18 nm on average. The total length of an associative chain, including the spacer and the sticker, is estimated from the molecular characteristics of the polymers and is of the order of 5 nm [43]. We deduce that loops where two stickers belonging to the same particle are connected together are very unlikely.

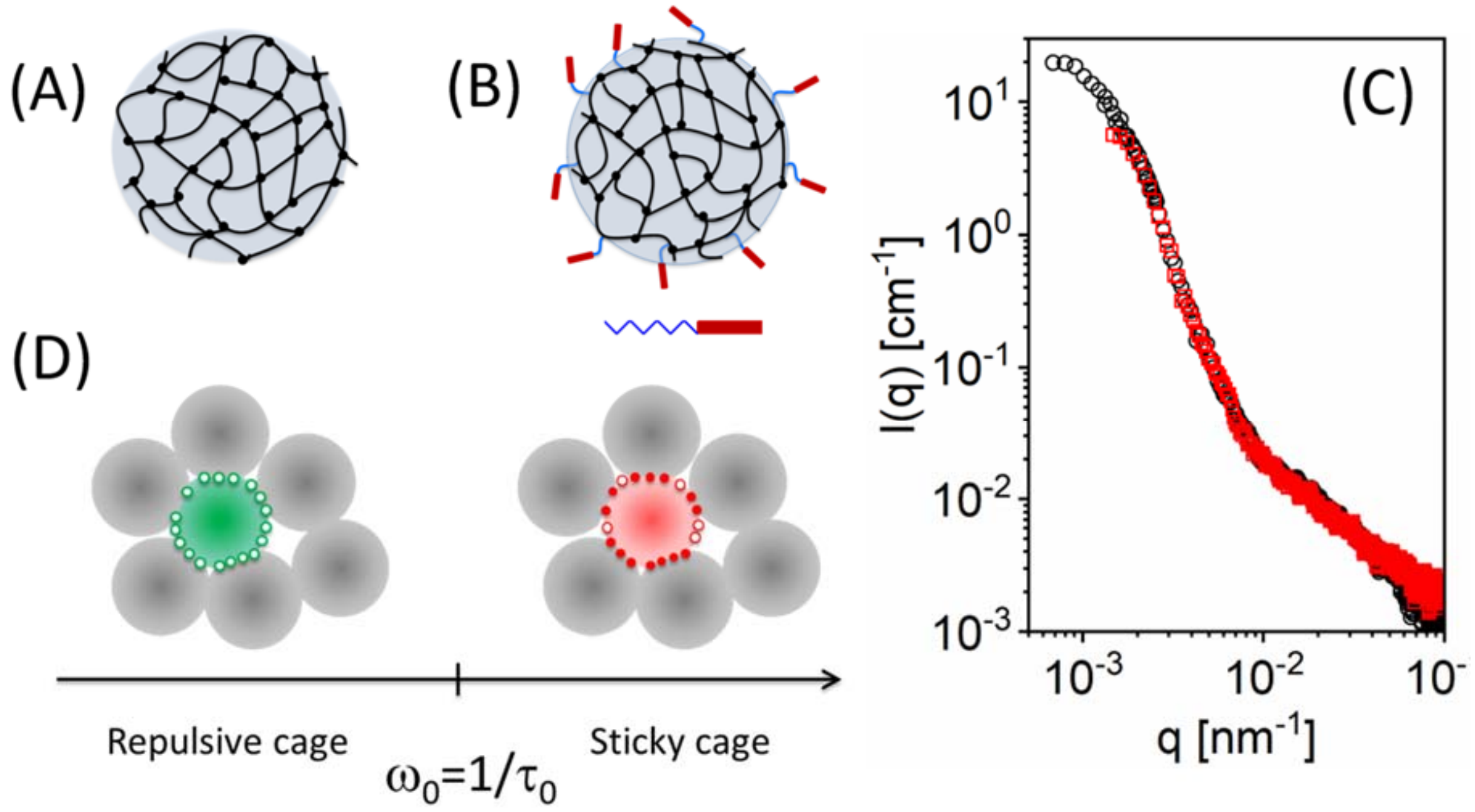

**FIG. 1.** Schematics of a repulsive (A) and associative (B) microgel. Associative microgels bear short chains made of a hydrophilic polyoxyethylene spacer -$(CH_2CH_2O)_{25}$- terminated by a linear alkyl chain -$C_{22}H_{45}$. (C): form factors of repulsive (black) and associative microgels (red); (D): schematic of the microstructure at rest of an associative SPG for different frequencies of observation. At high frequency, most stickers are associated (red disks) and particles (red color) are stuck in adhesive cages; at low frequencies, stickers are dissociated (green circles) and particles are trapped in repulsive cages PG is recovered (green color); the relaxation frequency $\omega_0$ is obtained from linear viscoelasticity measurements (Fig. 3A).

**B. Rheological states of suspensions and jamming transition**

We use linear and nonlinear rheology to characterize the rheological states of the suspensions and the jamming transition when the concentration varies. We extend the procedure developed earlier for suspensions of repulsive microgels to associative microgels [4]. We first measure the variations of the storage and loss moduli, $G'(\omega)$ and $G''(\omega)$, versus the angular frequency for a series of suspensions at increasing concentrations (Fig. 2A). When the polymer concentration is smaller than $C = 3.5$ mg.g$^{-1}$, the loss modulus is larger than the storage at low frequency with $G''(\omega) \sim \omega$ and $G'(\omega) \sim \omega^2$, indicating that the viscoelastic moduli reach terminal relaxation. At $C = 3.5$ mg.g$^{-1}$, the storage modulus is larger than the loss modulus down to $\omega \cong 10^{-3}$ rad.s$^{-1}$ where it crosses the loss modulus, signaling that terminal relaxation is approached but not reached within the experimental frequency window. At $C = 4.0$ mg.g$^{-1}$, the suspensions have a nearly frequency-independent storage modulus larger than the loss modulus over the entire accessible frequency window showing that they are solid-like.

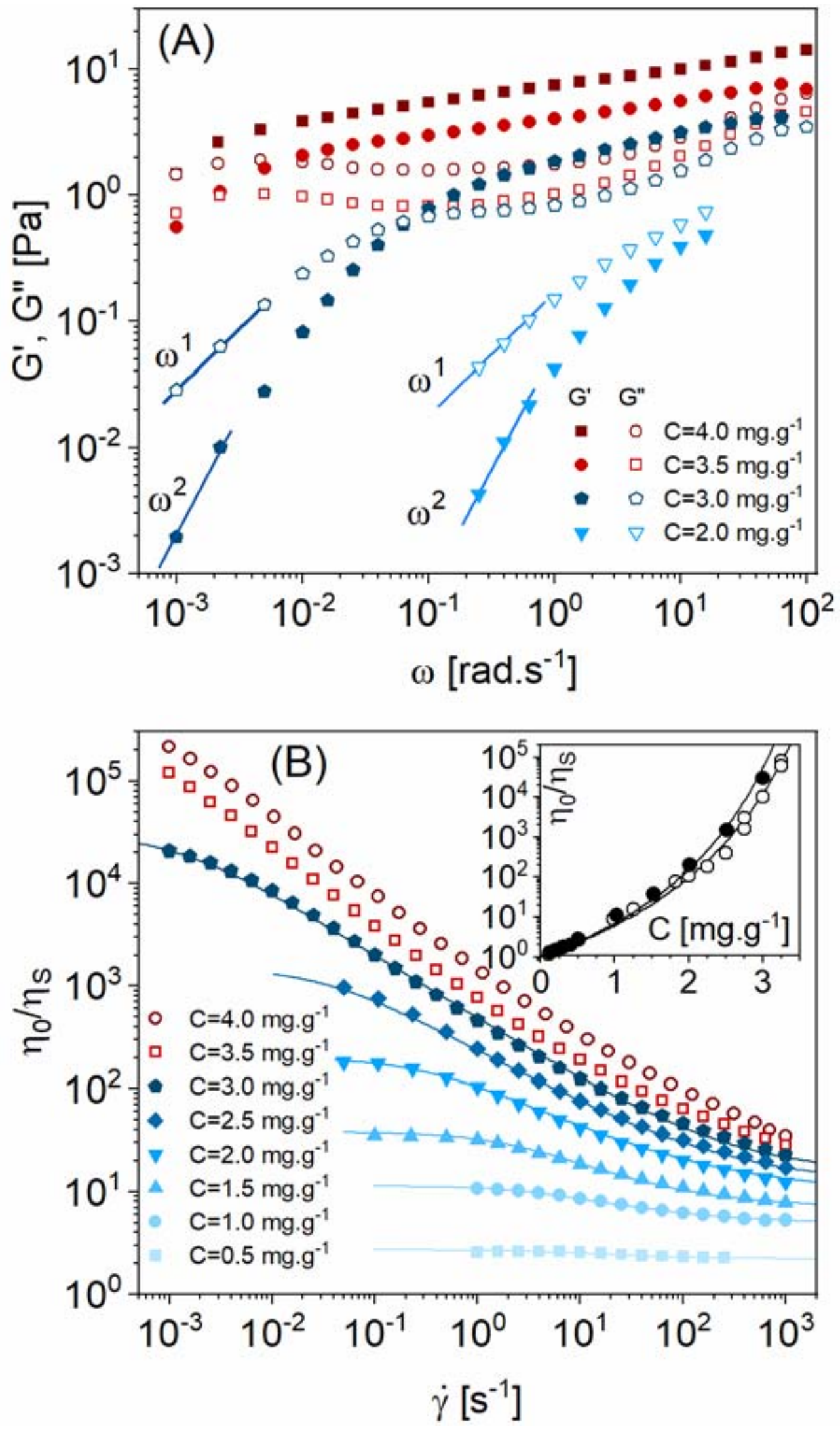

**FIG. 2** Determination of the liquid-solid transition of associative microgel suspensions. (A): variations of the storage (full symbols) and loss modulus (open symbols) for increasing polymer concentrations. (B) variations of the viscosity for increasing polymer concentrations; the continuous lines are fits of the experimental data to the Carreau-Yasuda model (Eq. 1). The inset shows the sharp increase of the low-shear viscosity with the polymer concentration for both the associative (full symbols) and repulsive (open symbols) microgels; the lines represent the best fits of the experimental data to the Doolitle equation.

We also measure the variations of the viscosity with the shear rate when the polymer concentration is increased. Figure 2B shows that at low concentration, the viscosity exhibits a Newtonian plateau at low shear rates. The low-shear plateau is followed by shear-thinning at intermediate shear rates and a second plateau at high shear rates. The experimental viscosity curves are well fitted to the Carreau-Yasuda equation (see Section II.C). The values of the low-shear viscosity are plotted in the inset of Fig. 2B. $\eta_0$ increases by several orders of magnitude and becomes inaccessible beyond $C \cong 3.5$ mg.g$^{-1}$, which is where the longest relaxation time falls outside the observation window. The viscosity data are well described by the following empirical form inspired from the Doolittle equation for hard sphere colloidal suspensions [44]:

$$\eta_0 \cong \eta_S \exp\left(\frac{AC}{C_m - C}\right). \tag{6}$$

$A \cong 8.2 \pm 0.2$ is a constant and $C_m \cong 5.3 \pm 0.2$ mg.g$^{-1}$ is the concentration where the low-shear viscosity diverges.

For the sake of comparison, we have performed the same analysis for repulsive microgel suspensions. The variations of the viscoelastic spectra and the shear viscosity versus the concentration are presented in Fig. S1 in supplementary material. A close examination of the viscoelastic spectra (Fig. S1A) reveals that the onset of terminal relaxation of repulsive microgel suspensions occurs at higher frequency than for associative microgel suspensions at the same concentration. From the shear viscosity of the repulsive microgel suspensions (Fig. S1B), we extract low-shear viscosity data, which are plotted in the inset of Fig. 2B. They can also be fitted to the Doolittle equation with $A \cong 8.5 \pm 0.2$ and $C_m \cong 5.7 \pm 0.2$ mg.g$^{-1}$. These results demonstrate that both types of microgel suspensions exhibit a liquid-solid transition, that we associate with the jamming transition. Below the jamming transition, they behave like non-Newtonian viscous dispersions and above it they are dynamically arrested. The variations of the low-shear viscosities are comparable but the jamming concentration of associative microgel suspensions is slightly shifted to a lower concentration. This shows that the interactions introduced by the alkyl group are short-ranged and do not significantly alter the behavior of the suspensions below the jamming concentration. Above the jamming transition, the microgels interact through contacting facets and form amorphous materials which behave like weak solids. Interactions between repulsive microgels in contact are dominated by elastic repulsive forces that can be described by a Hertzian potential [7] whereas interactions between hydrophobically-modified microgels involve a combination of repulsion and short-ranged hydrophobic associations.

### C. Linear viscoelasticity properties of Associative Soft Particle Glasses

The variations versus the applied frequency of the storage and loss moduli at small deformation of repulsive and associative SPGs at the same concentration $C = 15\,\mathrm{mg.g^{-1}}$ are shown in Fig. 3A. The storage modulus of the associative SPG exhibits two plateaus at high and low frequency, $G_+$ and $G_-$ respectively, which are separated by an inflection point at the intermediate frequency $\omega_0$. The loss modulus is lower than the storage modulus and has a broad maximum centered around $\omega_0$. The presence of this maximum signals the existence of dissipation processes at frequencies of the order of $\omega_0$. By contrast, the storage modulus $G'(\omega)$ of the repulsive SPG is weakly dependent on the frequency and the loss modulus $G''(\omega)$ exhibits a shallow minimum at an intermediate frequency $\omega_m$. At very low frequency the storage moduli of the repulsive and associative SPG glasses are comparable. The difference between the viscoelastic properties of associative and non-associative microgels highlights the importance of the functionalization of the microgels with associative groups.

We model the experimental viscoelastic spectra in Fig. 3 using the Havriliak-Negami equation as explained in Section II.C. It describes well the variations of *G*' over the entire range of frequency, and *G*'' around $\omega_0$ but not at low and high frequency, due to non-affine rearrangements [45] and elastohydrodynamic lubrication [46], respectively. It is interesting to apply the same analysis successfully to the other concentrations which are presented in Fig. S2 in supplementary material. The intermediate relaxation does not exist below $C = 6\,mg.g^{-1}$, which is close to the jamming transition previously determined from the divergence of the low-shear viscosity, further supporting that associations play a significant role only when the microgels are in contact. For 6 mg.g$^{-1}$≤$C$≤20 mg.g$^{-1}$, the parameter $\alpha$ is of the order of 0.47±0.02 showing that the relaxation at $\omega_0$ is broad but insensitive to the concentration.

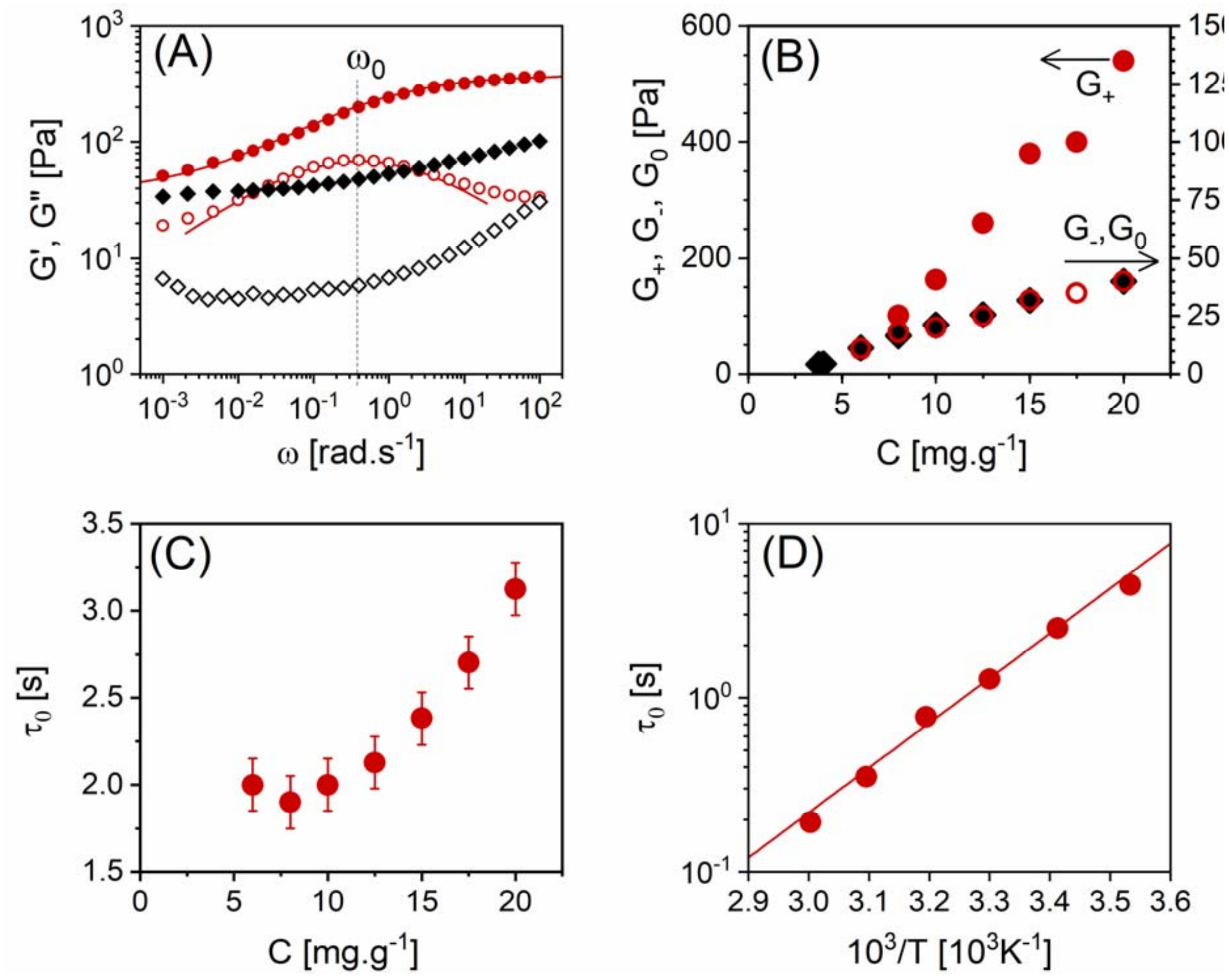

**FIG. 3** Linear viscoelastic properties of jammed microgel suspensions. (A): storage (●,◆ ) and loss moduli (○,◇) versus frequency for associative (●,○) and repulsive (◆,◇) microgels at $C$ = 15 mg.g$^{-1}$; the lines are fits to the Havrilak-Negami equation as explained in the text. (B): low (◆, ○; right axis) and high frequency (●; left axis) storage moduli versus concentration for associative (●,○) and repulsive microgels (◆); (C): relaxation time $\tau_0 = 1/\omega_0$ versus concentration; (D): Arrhenius plot of the relaxation time $\tau_0$ for $C$ = 15 mg.g$^{-1}$; the equation of the fitting line is: $\tau_0 = \tau^* e^{E/RT}$ with $\tau^*$ = 4 ns and $E$ = 49 kJ.mol$^{-1}$ ($R$ is the perfect gas constant).

The variations of $G_+$ and $G_-$ with concentration are plotted in Fig. 3B together with the moduli $G_0$ of the repulsive SPGs, which are taken at the minimum $\omega_m$ of the loss modulus. $G_-$ and $G_0$ coincide over the entire range of concentration investigated, as expected from Fig. 3A. Thus, the elastic behavior of associative and non-associative SPGs *at low frequency* has the same physical origin, i.e. the repulsive forces of elastic origin that the microgels exert on one another through their contacting facets [47]. The *high frequency moduli* $G_+$ are much larger than $G_- \cong G_0$, because of the additional contribution of the attractive interactions. In conclusion, associations play a negligible role at low frequencies but dominate the high frequency behavior of associative microgels.

Let us examine the properties of the characteristic relaxation time $\tau_0 = 1/\omega_0$. Figure 3C shows that $\tau_0$ weakly increases with the concentration. We have investigated the variations of $\tau_0$ within the range of temperature between 10 °C and 60 °C. Figure 3D is an Arrhenius plot of the values of $\tau_0$ as a function of $1000/T$. The data points fall on a straight line corresponding to $\tau_0 = \tau^* \exp\left(E_m/RT\right)$ where $\tau^*$ = 4 ns is a microscopic time and $E_m = 49\ \mathrm{kJ.mol^{-1}}$ is an activation energy. This activation energy corresponds to an energy of about 1 $k_BT$ per methylene group as already found in telechelic polymers [48]. Similar results are obtained for the other concentrations with comparable values of the activation energy. This demonstrates that the relaxation of the viscoelastic properties at intermediate frequency is driven by a thermally activated process and can be tuned by temperature.

We now return to the jammed structure represented in Fig. 1D to provide a microscopic interpretation of these results. Each microgel is trapped in a cage formed by several surrounding neighbors. Since the microgels are deformable, they form facets at contact through which they experience net repulsive forces of elastic origin and short-range attractive forces due to hydrophobic associations between stickers from both sides of the facets. These associations have a given lifetime and a necessary condition to relax stress is that stickers bridging contacting microgels detach. At high frequency, the lifetime of the associations is longer than the characteristic time of the applied deformation so that the stickers remain associated: the microgels are trapped in adhesive cages maintained by hydrophobic interactions. At low frequencies, all the stickers connecting microgels to their neighbors can dissociate and relax stress so that the microgels feel repulsive elastic forces. The characteristic relaxation frequency $\omega_0$ thus marks the transition between an associative glassy state where microgels are bonded together through adhesive forces and trapped in sticky cages, and a glassy state dominated by repulsive forces of elastic origin.

A pending question concerns the relation between the dynamics of individual stickers and the mechanical relaxation at frequency $\omega_0$. It has been argued that the mechanical relaxation of suspensions of particles connected by associative polymers generally involve several mechanisms [49]: the relaxation of the polymeric chains connecting the particles, the formation of loops on the surface of the same particles, the rearrangement of particles, and the breakup of associations bridging distinct particles. The first three mechanisms are not relevant here: the polymer chains are short and relax at high frequency unlike in associative polymers; particles are jammed and their relaxation takes place at very low frequency, if any; loops are not expected to form for the steric reasons given above. We are thus left with the dissociation of stickers as the dominant relaxation mode. This problem has motivated a micromechanical model that investigates the linear viscoelastic properties of jammed suspensions made of hydrophobically modified soft particles similar to ours [50]. The model is first constructed for horizontal surfaces decorated with stickers and then generalized to soft particle suspensions arranged in a cubic lattice. The storage modulus exhibits two plateaus as found experimentally, the low-frequency plateau corresponding to the plateau modulus computed for repulsive forces only. The loss modulus exhibits a local secondary relaxation at a frequency that corresponds to the lifetime of the bonds formed by the associated stickers through an $O(1)$ factor. In conclusion, the mechanical relaxation in associative SPGs is directly related to the molecular kinetics controlling the connectivity. The frequency at which the loss modulus exhibits a local maximum and the storage modulus an inflection that gives access to the dissociation time of the stickers. Interestingly, the same relation between the microscopic dynamics and the macroscopic viscoelasticity exists in telechelic polymer hydrogels [51].

**D. Oscillatory Yielding of Associative Soft Particle Glasses**

The dynamic nature of the associations has important consequences on the yielding properties of associative SPGs. In Fig. 4, we plot the variations of the storage modulus, loss modulus, and shear stress amplitude versus the strain amplitude, which are measured in Large Amplitude Oscillatory Shear Experiments, for three angular frequencies smaller, equal, and larger than the characteristic frequency $\omega_0$, respectively.

*For* $\omega << \omega_0$, the suspension has the same behavior as the repulsive SPG at the same concentration and for the same angular frequency (compare Fig. 4A and Fig. S3 in supplementary material. At low strain amplitudes, both moduli are independent of the strain amplitude with $G' >> G''$. The shear stress increases linearly with the strain amplitude, showing that the response is a linear function of the deformation. At large amplitude, the loss modulus exceeds the storage modulus, indicating that the suspension has yielded. The shear stress follows a power law represented by a straight line in double logarithmic coordinates. The intersection between the linear variation at low strain and the power law at high strain defines a unique yield point characterized by a yield strain of the order of 0.1, which corresponds to the yield strain of repulsive SPGs. In conclusion, at low frequencies, associative SPGs exhibit simple yielding associated to cage breaking and large scale rearrangements.

*For* $\omega \cong \omega_0$ (Fig. 4B), the stress amplitude first varies linearly, then becomes sublinear above the strain amplitude $\gamma_y^{(1)} \cong 0.2$, goes through an inflection point and finally reaches another sublinear variation above a second yield strain $\gamma_y^{(2)} \cong 1.5$. This two-step yielding is associated with a first decrease of the storage modulus above $\gamma_y^{(1)}$ followed by a sharp drop after $\gamma_y^{(2)}$. The loss modulus exhibits a clear maximum and crosses the storage modulus near $\gamma_y^{(2)}$, where complete fluidization takes place.

*When* $\omega >> \omega_0$ (Fig. 4C), the stress exhibits a sharp break corresponding to the maximum of the loss modulus at the yield strain $\gamma_y^{(2)} = 1.5$. The loss modulus starts increasing at a strain of about 0.2 which is reminiscent of the first yielding process detected at lower frequency. This double-yielding process found for associative SPGs at high frequency resembles that found in attractive colloidal glasses consisting of mixtures of hard sphere colloids and non-adsorbing polymers responsible for depletion interactions [52], or colloidal gels formed in the presence of attractive interactions [53]. We attribute the first yield point $(\gamma_y^{(1)} \approx 0.2)$ to the local disruption of associations and in-cage rearrangements and the second yield point at large strain ($\gamma_y^{(2)} \approx 1.5$) to cage breaking and large scale rearrangements. In conclusion, the presence of dynamic attractive interactions affects the yielding behavior of associative SPGs: yielding can be simple as in conventional SPGs or double as in attractive glasses and gels, depending on the value of the frequency with respect to the relaxation frequency of associations.

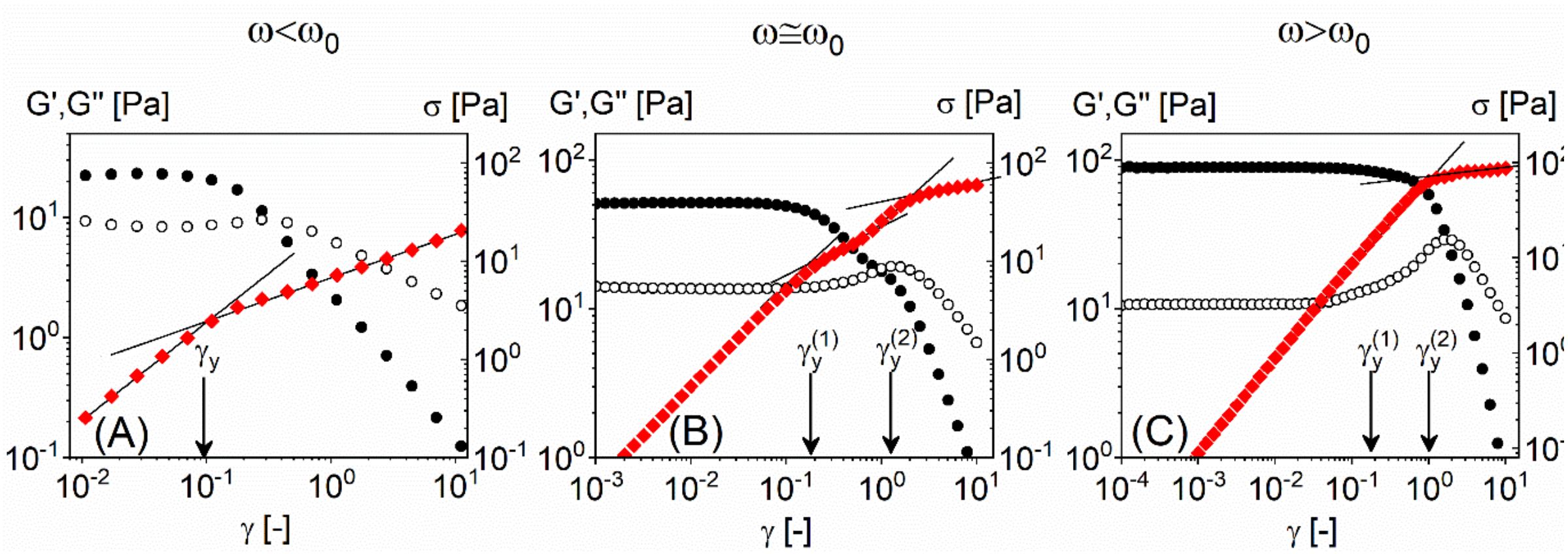

**FIG. 4** Variations of the storage modulus (●), loss modulus (○), and stress amplitude (◆) versus the strain amplitude in LAOS experiments at $\omega = 0.02$ rad.s$^{-1}$ (A), $\omega = 0.6$ rad.s$^{-1}$ (B), and $\omega = 25$ rad.s$^{-1}$ (C) for an associative microgel suspension at $C = 8$ mg.g$^{-1}$. The black lines in (A) and (C) are the best fits to a linear ( $\gamma < \gamma_y$ ) and power law function ( $\gamma > \gamma_y$ ); they intersect at the yield strain. The yield strain values in (B) are estimated from the intersection of the black lines that represent the local slopes of the stress curve.

## E. Continuous Flow of Associative Soft Particle Glasses

In this section, we analyze the flow curves giving the variations of the stress versus the apparent shear-rate applied or measured by the rheometer using controlled-rate and controlled-stress measurmements. Figure 5A shows the flow curves of repulsive and associative microgel glasses at the same polymer concentration, $C = 15\ \mathrm{mg.g^{-1}}$. The flow curve of the repulsive microgel SPG has the characteristic shape that is expected: a slowly decreasing plateau at low shear rates and a power law increase at high shear rates. The velocity profiles measured using μ-PIV are presented in Fig. S4 in supplementary material. They are perfectly linear without any evidence of wall slip, as already reported for glass-like materials with repulsive interactions [54]. In Fig. 5A, the apparent and true shear rates determined from the local velocity profiles coincide over the entire range of shear rate investigated.

In *controlled-strain experiments*, the flow curve of the associative microgel SPG exhibits three different regimes limited by two characteristic shear rates: $\dot{\gamma}_a^*$ where the stress starts to level off to a plateau and $\dot{\gamma}_a^{**}$ above which the flow curves of the associative and repulsive glasses coincide. For $\dot{\gamma}_a < \dot{\gamma}_a^*$ (called regime I in the following), the stress is significantly larger than for the repulsive microgel glass and increases monotonically. For $\dot{\gamma}_a^* < \dot{\gamma}_a < \dot{\gamma}_a^{**}$ (regime II), the stress reaches a constant plateau. For $\dot{\gamma}_a > \dot{\gamma}_a^{**}$ (regime III), the flow curves of the associative and repulsive SPGs are superimposed suggesting that the associations no longer contribute to the shear stress. In *controlled-stress experiments,* i.e., in creep experiments, the intermediate plateau is not accessible; the apparent shear rate measured at steady state falls either in branch I or III after some transient period (Fig. S5 in supplementary material.

Figure 5(B-C) represents the velocity profiles of the associative SPG, measured in controlled strain experiments, for values of the apparent shear rate spanning regimes I and II.

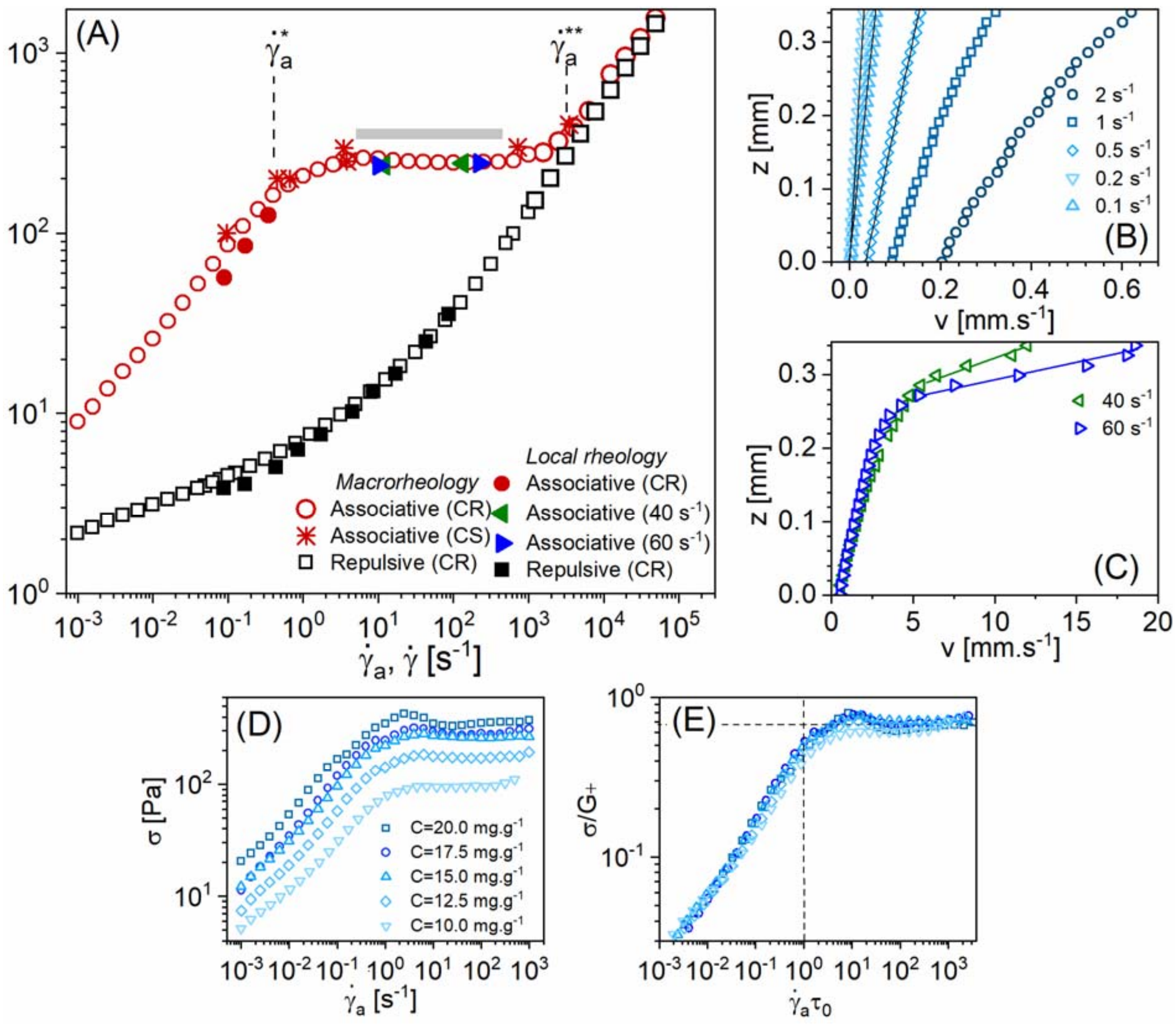

**FIG. 5** (A): Flow curves of associative (red) and repulsive (black) microgel glasses ($C = 15\ \mathrm{mg.g^{-1}}$) measured using macroscopic rheology in rate-controlled mode (open symbols, CR) and stress-controlled mode (star symbols, CS), and local rheology (full symbols) as explained in the main text. The triangles are obtained from the shear-banded profiles shown in (C). The grey bar indicates the range of shear rates which is not accessible using creep experiments. $\dot{\gamma}_a^{*}$ designates the shear rate where the stress starts to level off; the flow curves for the associative and repulsive microgel glasses meet at $\dot{\gamma}_a^{**}$ (B) and (C): velocity profiles measured with μPIV for different values of the apparent shear rate. The straight lines are linear fits to the velocity profiles giving access to the true shear rates that are reported in (A). (D): Flow curves of associative microgel glasses with increasing concentrations measured in rate-controlled mode (open symbols. (E): Master curve obtained from the data shown in (D) when the apparent shear rate is scaled by the viscoelastic relaxation frequency $\omega_0 = 1/\tau_0$ and the stress by the high frequency modulus $G_+$ (see Fig. 1). The dashed vertical and horizontal lines represent the equations $\dot{\gamma}_a \tau_0 = 1$ and $\sigma = 0.7 G_+$.

In regime I ($\dot{\gamma}_a \leq \dot{\gamma}_a^*$), the velocity profiles are linear, wall slip is absent, and the true and apparent shear rates are in good agreement so that the local rheology data coincide with the macroscopic data (Fig. 5A). In regime II ($\dot{\gamma}_a^* \leq \dot{\gamma}_a \leq \dot{\gamma}_a^{**}$), the velocity profiles change significantly. At shear rates just above $\dot{\gamma}_a^*$, the apparent velocity does not vanish at the fixed bottom plate, signaling the occurrence of wall slip; slip is important since for $\dot{\gamma}_a = 2\ \mathrm{s}^{-1}$, the slip velocity at the bottom plate is about 1/3 of the applied velocity. Moreover, the velocity profiles are not linear and it is no longer possible to define unambiguously a unique shear rate. At larger applied apparent shear rates corresponding to the stress plateau (Fig. 5C), wall slip vanishes; for instance, when $\dot{\gamma}_a = 40\ \mathrm{s}^{-1}$, the slip velocity at the bottom plate is only 3% of the applied velocity. The velocity profiles are composed of two bands flowing at two different velocities, the slow and fast bands being near the fixed and mobile walls, respectively. In conclusion, the flow of the associative SPG in regime II is non-homogeneous with an interplay between wall slip and shear-banding. Slip signals the onset of the flow heterogeneities but it becomes negligible when shear-banding is fully established.

To test the generality of these results, we have measured the flow curves of associative SPGs at concentrations ranging from $C$ = 10 mg.g$^{-1}$ to 20 mg.g$^{-1}$. The variations of the stress in regimes I and II, which are reported in Fig. 5D, are qualitatively similar but some quantitative differences can be noted. When the concentration increases, we observe that the onset of the stress plateau is shifted to lower apparent shear rates and the value of the stress plateau increases. In Fig. 5E we replot the five sets of data using dimensionless variables, scaling the apparent shear rate by the relaxation time determined from the linear viscoelastic properties and the stress by the high frequency modulus. We observe a reasonable collapse of the different flow curves onto a unique master curve. This master curve shows that the onset of slip and shear-banding occurs when the shear rate becomes comparable to the relaxation time of the associations. The stress plateau is determined by the high-frequency modulus according to $\sigma \cong 0.7G_+$.

The existence of shear-banded flow associated with a constant stress plateau at intermediate shear rates, has been reported for various types of soft materials [55], such as wormlike micellar solutions [56], micellar cubic phases [57], colloidal rod glasses [58], associative polymer networks [59]. In these systems, the stress versus the shear rate has a characteristic non-monotonic S shape; shear banding is a mechanical instability called gradient-banding instability that is driven by the sharp decrease of the stress with the shear rate at the onset of banding. For wormlike micellar [56] and telechelic polymer solutions [60], master curves similar to that in Fig. 5E can be built by scaling the stress and the shear rate by the plateau modulus and the relaxation time measured in the linear viscoelastic limit. Despite this analogy, there is a fundamental difference between micellar or polymeric solutions and associative SPGs: solutions are viscoelastic materials whereas associative SPGs are yield stress materials that are solid-like at rest. Generally, homogeneous flow is not possible in yield stress fluids with attractive interactions below a finite apparent shear rate [61]. In the following, we propose a microscopic model that accounts for the stress in associative SPGs under flow and clarifies the origin of shear-banding.

## F. Microscopic model

### *1. Model description*

We assume that the total stress can be decomposed into two contributions that are independent and additive: $\sigma = \sigma_e + \sigma_a$. $\sigma_e$ is associated with the rearrangement of the microgel particles and $\sigma_a$ comes from the dynamic associations that connect the particles. In a repulsive SPG, only the term $\sigma_e$ is present; it originates from the elastic distortions - detectable through the pair distribution function- which take place during the out-of-cage rearrangements induced by shear [7]. In a suspension of associative particles, there is an extra contribution, $\sigma_a$, which comes from the existence of associations. Let $N_a$ be the number of stickers that form bridges connecting a microgel to its neighbors; these associated stickers coexist with $N_d$ dissociated stickers that are pendant and do not connect microgels (Fig. 1D). The total number $N_0 = N_a + N_d$ depends on the number of microgels in contact and is a function of concentration; it is expected to be smaller than the total number of stickers per microgel. Each associated sticker experiences a force $f$ so that the stress $\sigma_a$ can be expressed as:

$$\sigma_a = f\frac{N_a}{4\pi R^2} \tag{7}$$

To compute $f$, we neglect the orientation of the stickers with respect to the shear direction and we assume that the total force $f_p$ acting on a particle is equally distributed over $N_a$ stickers so that $f = f_p / N_a$. Since the modulus of individual microgels is larger than the modulus of the jammed suspensions [4,45], we treat microgels as non-deformable particles of radius $R$ moving in an effective medium of elastic modulus $G$, for which $f_p = 6\pi G R^2 \gamma$ [62], and:

$$f = 6\pi G R^2 \gamma / N_a \tag{8}$$

where $\gamma$ is the local strain. The force $f$ stretches the stickers connecting the particles, speeds up dissociations, and reduces the lifetime of associations, $\tau_d$, with respect to the lifetime at rest, $\tau_d^0$. We assume that $\tau_d$ results from a shear-activated process and we use the Tanaka-Edwards formalism, which has been successfully applied to compute the stress in solutions of telechelic polymers [63-65]:

$$\tau_d = \tau_d^0 \exp\left(-\frac{f}{f_0}\right), \tag{9}$$

where $f_0 = k_B T/\delta$. The full problem can be solved numerically but it is possible to get an approximate solution analytically, by linearizing Eq (9): $\tau_d/\tau_d^0 = 1 - f/f_0$, with $\frac{f}{f_0} = \frac{6\pi G R^2 \delta}{N_a k_B T}\gamma$. The elastic modulus $G$ is dominated by the contribution of the associations (Fig. 3B) and it can be estimated as $G = \nu k_B T$ where $\nu$ is the density of associations in the continuous phase. Setting $\nu \cong N_a/4\pi R^2 \delta$, we get $f/f_0 \cong \gamma$. The quantity $\gamma$ represents the total strain accumulated during the lifetime of associations and is thus $\gamma = \dot{\gamma}\tau_d$, $\dot{\gamma}$ being the applied shear rate so that:

$$\frac{\tau_d}{\tau_d^0} = \frac{1}{1+\dot{\gamma}\tau_d^0} \quad \text{and} \quad \frac{f}{f_0} = \frac{\dot{\gamma}\tau_d^0}{1+\dot{\gamma}\tau_d^0} \tag{10}$$

Finally, we need to express the number $N_a$ of associations in the stress equation (7). We describe the sticker dynamics by a reaction exchange between $N_a$ associated and $N_d$ dissociated stickers. At steady state, $N_a$ and $N_d$ obey the detailed balance equation:

$$k_a N_d = k_d N_a \ , \tag{11}$$

where $k_a$ is the rate of association and $k_d$ the rate of dissociation. $k_d$ is the inverse of the lifetime of associations: $k_d = 1/\tau_d$; it increases with the applied shear rate according to (9). Similarly $k_a$ is the inverse of the association time that a free sticker takes to become engaged in an association. We assume here that it is not affected by the flow and depends only on the local geometry: $k_a = 1/\tau_a^0$, where $\tau_a^0$ is the association time at rest. The ratio $\alpha = \tau_a^0/\tau_d^0$ of the association time and lifetime in the quiescent state is an adjustable parameter in the following. It is expected to be smaller than 1 since otherwise most stickers would be free. Finally, we obtain the following expression for the number of associated stickers per particle:

$$N_a = \frac{N_0}{1+\alpha\left(1+\dot{\gamma}\tau_d^0\right)} \tag{12}$$

## *2. Model predictions*

Using Eqs (7) - (11), we can derive the stress component $\sigma_a$ due to the dynamics of stickers. It is convenient to express $\sigma_a$ in the non-dimensional form:

$$\frac{\sigma_a}{\sigma_a^0} = \frac{\dot{\gamma}\tau_d^0}{1+\dot{\gamma}\tau_d^0}\frac{1}{1+\alpha(1+\dot{\gamma}\tau_d^0)}\ , \tag{13}$$

The prefactor $\sigma_a^0 = k_B T N_0 / 4\pi\delta R^2$ can be interpreted as the elastic modulus when all the stickers are associated so that $\sigma_a^0 \cong G_+$. In Fig. 6A, we represent the variations of $\sigma_a/\sigma_a^0$ as a function of $\dot{\gamma}\tau_d^0$ for different values of α. The stress is a non-monotonic function of the reduced shear rate. For $\dot{\gamma}\tau_d^0 << 1$, we have $\sigma_a/\sigma_a^0 \simeq \dot{\gamma}\tau_d^0/(1+\alpha)$ and the stress is an increasing function of the shear rate. For $\dot{\gamma}\tau_d^0 >> 1$, we have $\sigma_a/\sigma_a^0 \simeq 1/\left(1+\alpha\dot{\gamma}\tau_d^0\right)$ and the stress is a decreasing function of the shear rate. When α decreases from 1 to $10^{-2}$, i.e. when the lifetime of the association at rest increases, the position of the maximum, $\left(\dot{\gamma}\tau_d^0\right)_m = \sqrt{\alpha^{-1}+1}$ ($\alpha \neq 0$), shifts from about 1 to 10 and the amplitude from 0.2 to 0.8, respectively. Therefore the maximum stress is reached for a reduced shear rate $\dot{\gamma}\tau_d^0$ of order 1. $\dot{\gamma}\tau_d^0$ represents the Weissenberg number below which the sticker dynamics is dominated by Brownian effects and above which it is shear-activated. When the shear rate increases, the lifetime of the associations decreases up to the point where it becomes comparable to the characteristic time scale of the shearing motion.

The non-monotonic variation of $\sigma_a$ causes a decrease of the total stress $\sigma = \sigma_a + \sigma_e$ above some critical shear rate, since the term $\sigma_e$, which is associated with particle rearrangements, monotonically increases with the shear rate. For $\alpha \leq 1$, the variations of the reduced stress $\sigma / G_+$ as a function of the reduced shear rate $\dot{\gamma}\tau_d^0$ have the characteristic S-shape associated with a gradient-banding instability (see supplementary material. This is shown in Fig. 6B for $C$ =15mg.g$^{-1}$. The lifetime of the associations deduced from the linear viscoelastic data is $\tau_d^0 = 2.4$ s and the high frequency modulus is $G_+ = 360$ Pa . The stress measured for the purely repulsive SPG is well represented by $\sigma_e = 1.6 + 3.5\dot{\gamma}^{0.24} + 2.1\dot{\gamma}^{0.61}$. We have set $\alpha = 0.05$ for which the stress reaches a maximum value $\sigma_{\max} \simeq 0.7G_+$ at $\dot{\gamma}\tau_d^0 \simeq 4$. We distinguish three flow regimes as in Fig. 5(A-E). In regime I, the stress of the associative SPG is significantly higher than for the repulsive SPG because the association-dissociation dynamics of the stickers increases the dissipation. The lifetime of the associations decreases up to the point where it becomes comparable to the characteristic time scale of the shearing motion. In regime II, the actual stress decreases down to a value $\sigma_{\min}$ , which causes an instability where the flow must separate into two coexisting bands sheared at different shear rates. The apparent shear stress plateau may in principle lie anywhere between $\sigma_{\max}$ and $\sigma_{\min}$ depending on the existence of a specific selection rule [66]. Regime III is signaled by the upturn of the stress when the stress $\sigma_e$ associated with rearrangements takes over. The lifetime of the stickers $\tau_d$ becomes shorter than the association time $\tau_a \cong \tau_a^0$ , meaning that associations are irreversibly broken and do not have time to reform.

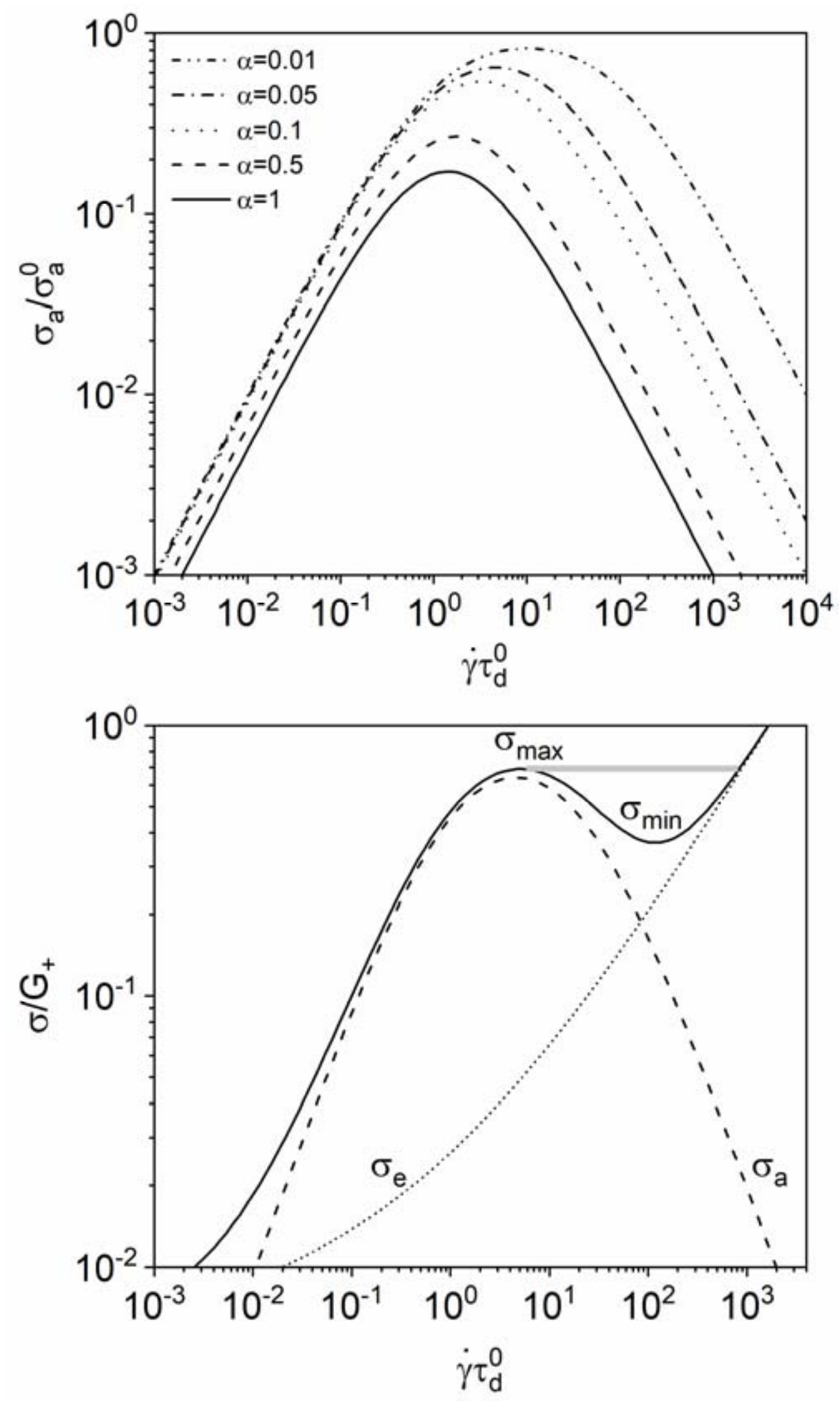

**FIG. 6** (A) Non monotonic variations of the stress due the sticker dynamics versus the reduced shear rate for different values of the parameter α, which represents the ratio of the association time to the lifetime in the quiescent state. (B) Schematic variations of the reduced stress ($\sigma = \sigma_a + \sigma_e$) versus the reduced shear rate (full line) for α=0.05; the dashed line represents the stress computed in the unstable region, the grey bar represents the stress plateau in the shear-banded region assuming that flow remains stable up to $\sigma_{\max}$.

## IV. Conclusions

Dynamic associations are responsible for remarkable time-dependent properties in jammed materials. In the jammed state, where particles are stuck in cages formed by their contacting neighbors at short time, the dynamics of the association is responsible for the thermally activated relaxation of the sticky cages at a frequency $\omega_0$ . The suspensions behave like repulsive glasses at long times ( $t > 1/\omega_0$ ) or small frequencies ( $\omega < \omega_0$ ) and colloidal gels at short time ( $t < 1/\omega_0$ ) or large frequencies ( $\omega > \omega_0$ ). The value of $\omega_0$ depends on temperature and on the composition of the associating groups; here we used alkyl groups comprising 22 carbon atoms but preliminary results obtained with other compositions and architectures or stickers indicate that it is possible to tune the relaxation frequency and adjust its value to specifications requested by applications [38].

The relaxation of in-cage associations not only controls the viscoelastic properties in the quiescent state but also the yielding properties and the flow behavior of the jammed suspensions. Whereas attractive SPG are expected to flow homogeneously when associations are not dynamical [30-33], we have shown experimentally and theoretically that shear-banding can occur when associations have an intrinsic lifetime that depends on the shear rate. The macroscopic rheology depends on the experimental protocol. In controlled strain experiments, a stress plateau where the flow comprises two bands moving at different velocities separates a low shear and a high shear region where the flow is homogeneous. In the low shear domain, associations bring an extra contribution to the stress associated with rearrangements, which can be interpreted in terms of local friction. It is interesting to note that the onset of shear-banding is signaled by strong wall slippage which disappears deeper inside the banded flow regime. Wall slip, which is absent in the purely repulsive SPGs, thus appears as an alternative mechanism to dissipate energy in inhomogeneous flow. At high shear rates, associations are disrupted and do not reform so that the stress coincides with that measured in purely repulsive suspensions. In controlled stress experiments, the shear-banded plateau does not exist and the flow curve suddenly jumps from the low-shear to the high shear branch. Such flow curves thus differ from the non-monotonic flow curves described for yield stress fluids, where no steady homogeneous flow is observed below a critical shear rate [35,36,61], reflecting different mechanisms.

We have proposed a microscopic model that accounts for the existence of shear banding in jammed colloidal suspensions with dynamical associations. Shear-banding appears to be a mechanical gradient-banding instability caused by the strong drop of the relaxation time, i.e. shear-thinning, when associations are pulled apart by shear forces. The shear rate at which the flow becomes non-homogeneous with wall slip and/or shear banding is set by the inverse of the relaxation time of the associations. In conclusion, the existence of dynamic bonds profoundly affects the dynamics of yield stress fluids both in the linear and nonlinear regimes. Our findings must be relevant in a wide range of soft materials. To go further and test the generality of our results, it would be interesting to implement other types of dynamic associations based on metal-ligand supramolecular chemistry [24], host-guest interactions [67], H-bonding [68], or reversible covalent chemistry [69].

## SUPPLEMENTARY MATERIAL

The supplementary material includes additional figures that further support and complement the results presented in this study: jamming transition of a microgel SPG with purely repulsive interactions (SM1), viscoelastic spectra of associative microgel SPGs at different concentrations (SM2), determination of the yield point of a repulsive microgel SPG (SM3), velocity profiles of a repulsive microgel SPG (SM4), controlled stress rheology of an associative SPG (SM5), nonmonotonic stress in associative SPGs (SM6).

## ACKNOWLEDGMENTS

The authors are indebted to Roger Bonnecaze and Fardin Khabaz for enlightening discussions about the rheology of SPG. MM has benefited from financial support from the European Training Network **COLLDENSE** (*Hybrid Colloidal Systems with Designed Response* - grant 642774.

## DATA AVAILABILITY

The data that support the findings of this study are available from the corresponding author upon reasonable request.

## AUTHOR DECLARATIONS

### Conflict of Interest

The authors have no conflicts to disclose

# Supplementary Material

## Shear-Banding Driven by Dynamic Associations

## in Soft Particle Glasses

Maddalena Mattiello[*], Paolo Edera[*], Jean-Marc Suau[**], Michel Cloitre[*]

[*] Molecular, Macromolecular Chemistry, and Materials, ESPCI Paris, PSL Research University, CNRS, 75005 Paris, France

[**] Arkema, ZI Lyon Nord, 35 Rue Ampère, 69730 Genay, France

Corresponding author: michel.cloitre@espci.psl.eu

## Table of content

**SM1. Determination of the jamming transition of a microgel SPG with purely repulsive interactions**

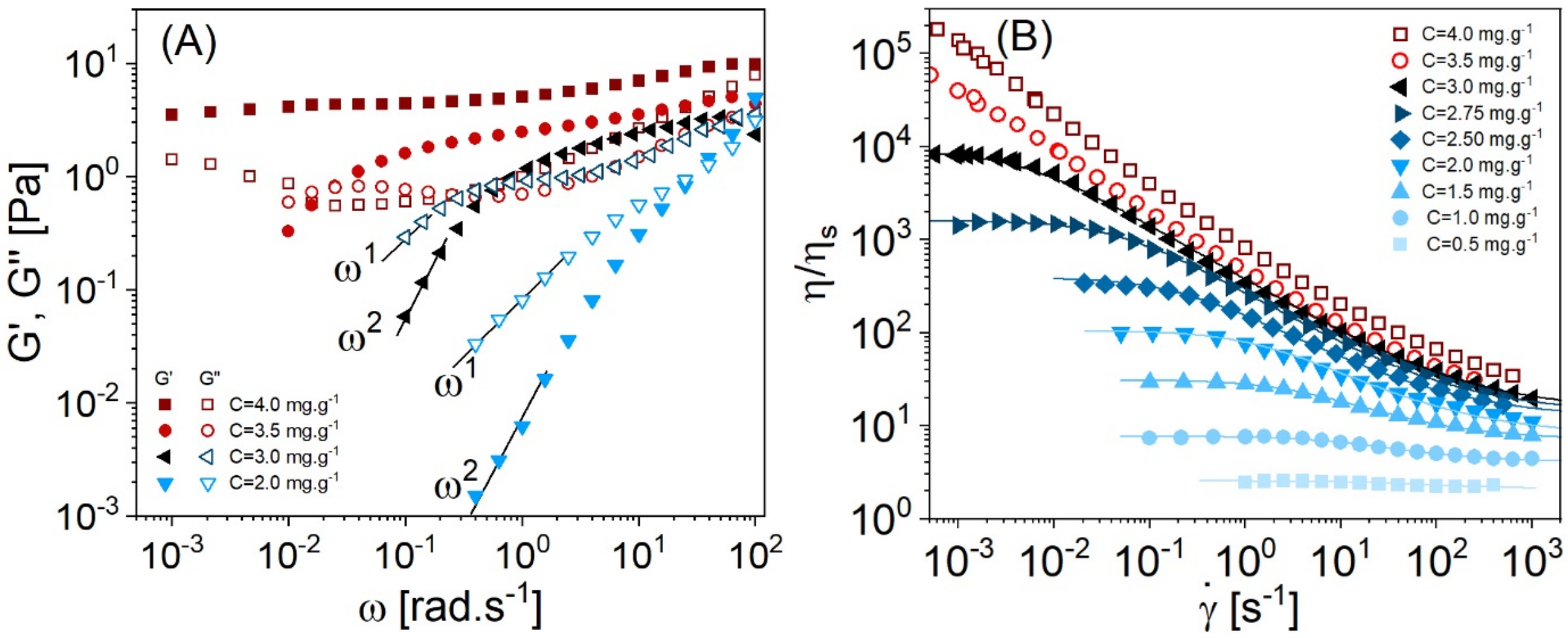

**FIG. S1** Determination of the liquid-solid transition of repulsive microgel suspensions. (A): variations of the storage (full symbols) and loss modulus (open symbols) for increasing polymer concentrations. (B) variations of the viscosity for increasing polymer concentrations; the continuous lines are fits of the experimental data to the Carreau-Yasuda model (Eq. 1 in the main text). Above the jamming concentration, terminal relaxation does not occur within the experimentally accessible time-scale window. We estimate that the jamming concentration is slightly higher than $C = 3.0 \text{ mg.g}^{-1}$.

**SM2. Viscoelastic spectra of associative microgel SPGs at different concentrations**

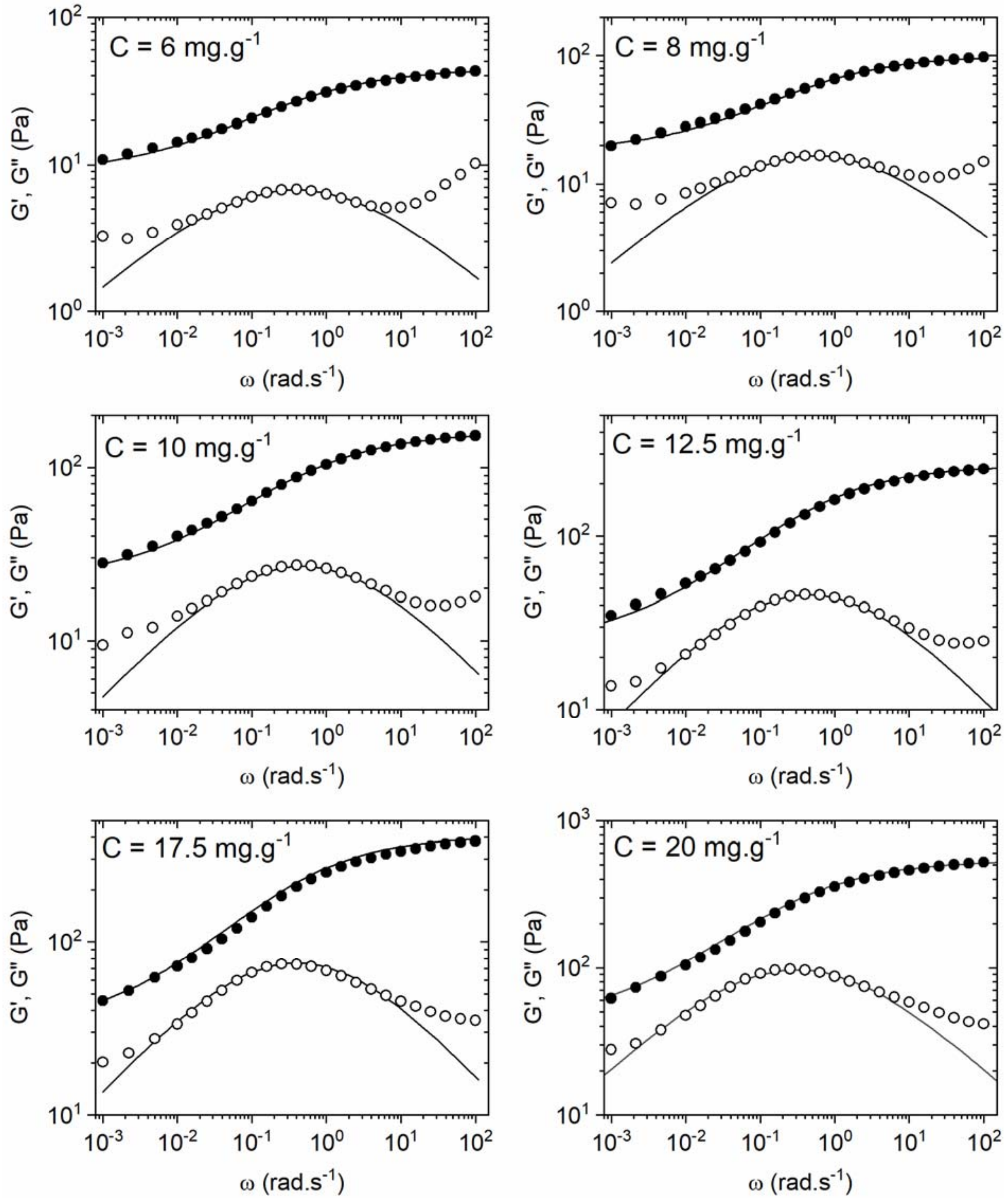

**FIG. S2** Storage (● ) and loss (○) moduli versus angular frequency for associative microgel SPGs at increasing concentrations. The lines are fits to the Havrilak-Negami equation as explained in the main text.

**SM3. Large Amplitude Shear Rheology and determination of the yield point of a repulsive microgel SPG.**

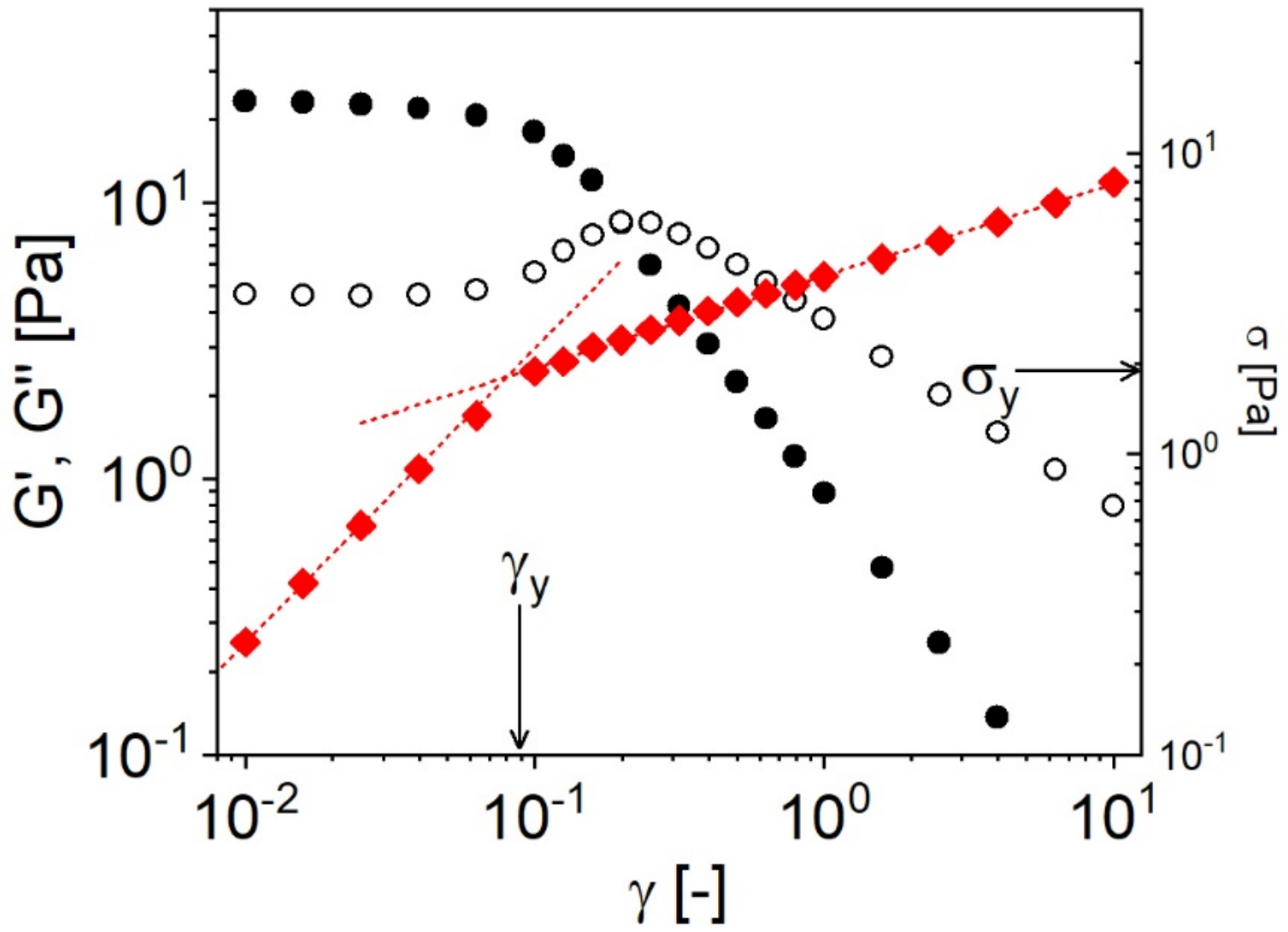

**FIG. S3** Variations of the storage modulus (●), loss modulus (O), and stress amplitude (◆) versus the strain amplitude in LAOS experiments at $\omega = 0.02$ rad.s$^{-1}$ for a repulsive microgel SPG at $C = 8$ mg.g$^{-1}$. The dotted lines are the best fits to a linear variation ($\gamma < \gamma_y$) and a power law variation ($\gamma > \gamma_y$); they intersect at the yield strain $\gamma_y$.

## SM4. Velocity profiles of a repulsive microgel SPG

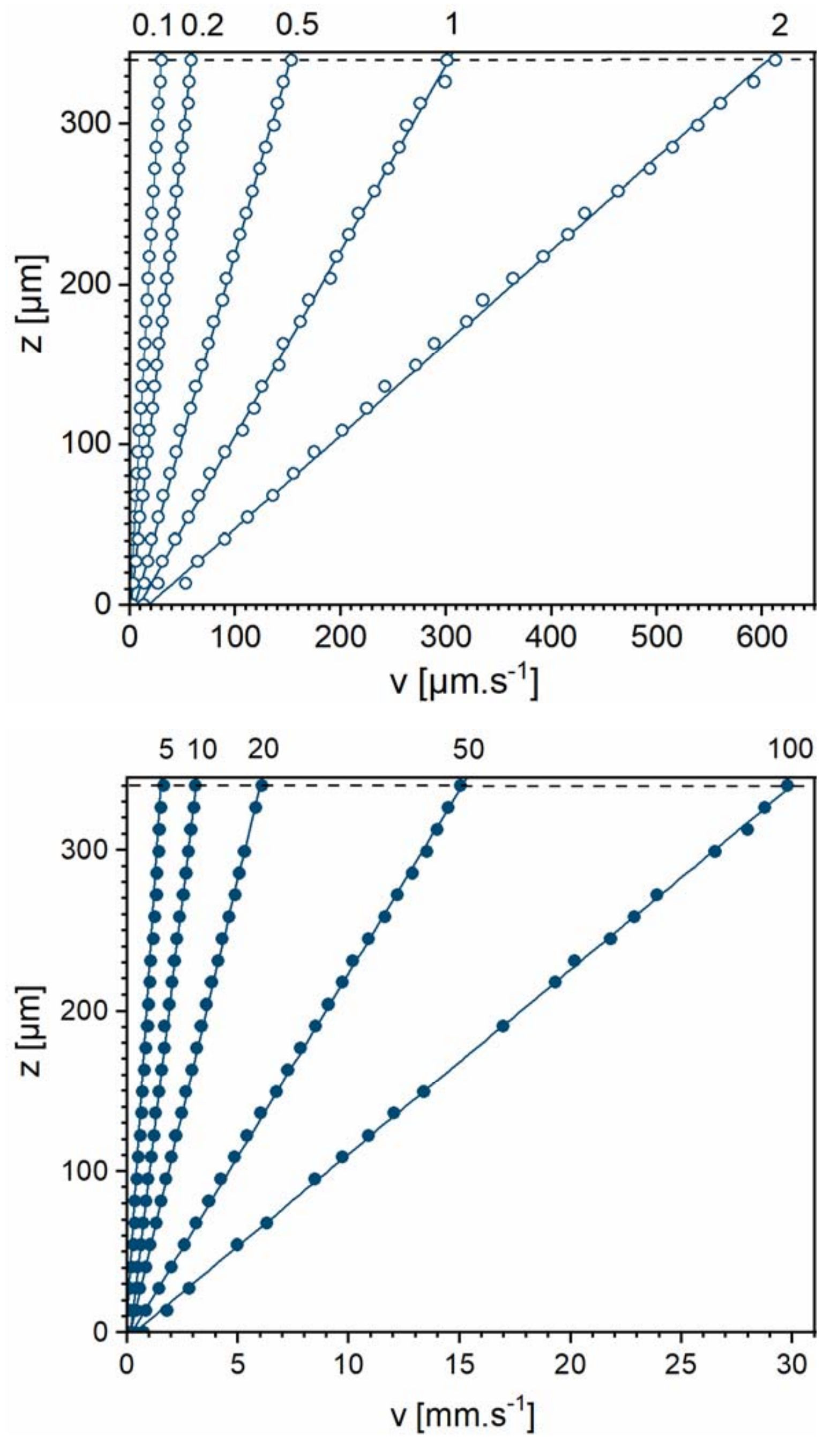

**FIG. S4** Velocity profiles of a repulsive SPG at $C$ = 15 mg.g$^{-1}$, measured using µPIV for increasing values of the apparent shear rate (from left to right and top to bottom). The straight lines are linear fits to the velocity profiles giving access to the true shear rates which are reported in Fig. 5A.

## SM5. Controlled stress rheology of associative SPGs

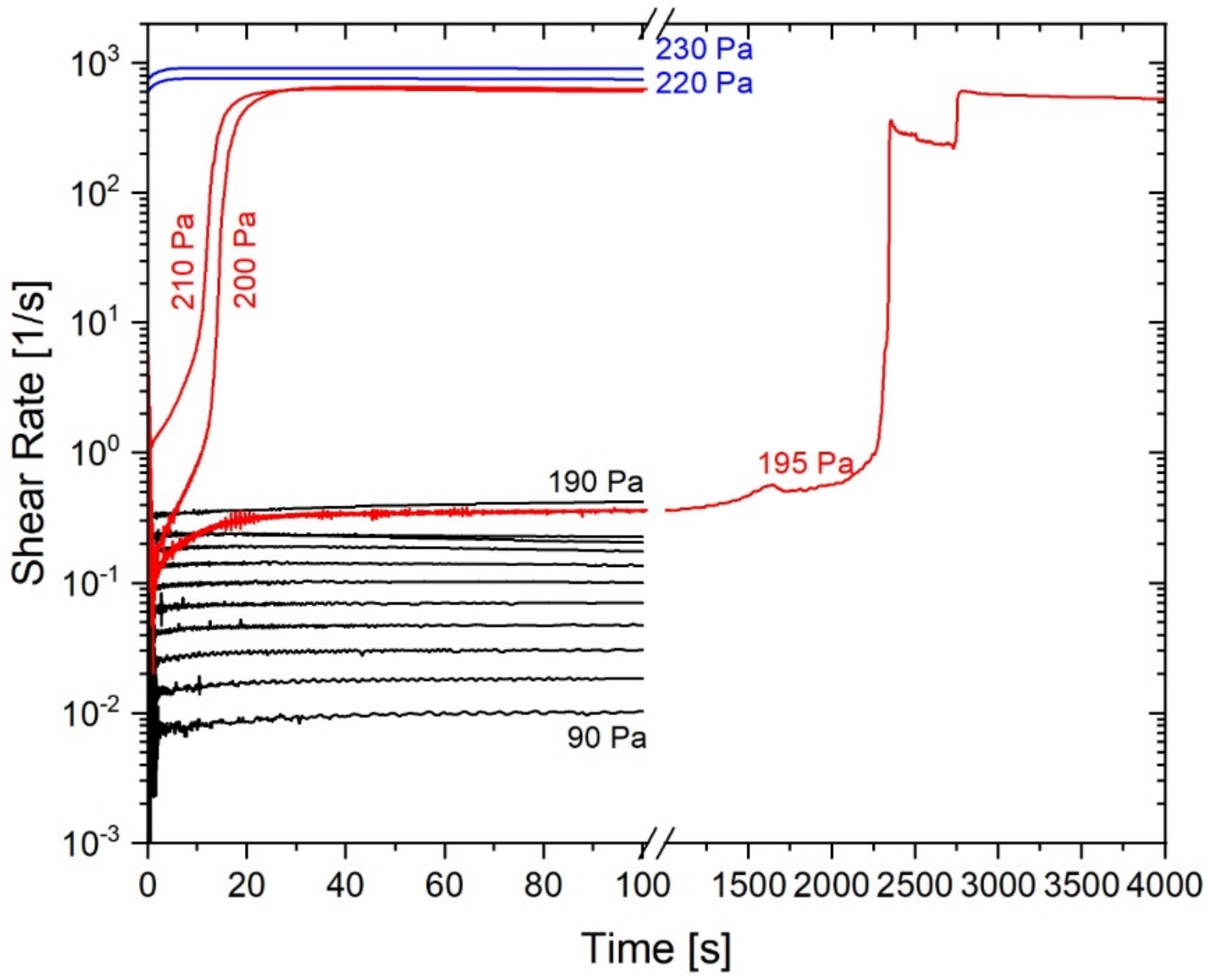

**FIG. S5** Creep properties of the associative SPG at $C$ = 15 mg.g$^{-1}$. Stress of increasing values are applied and the shear rate response is recorded until steady state is reached. Such measurements give evidence for two flow branches at low and high shear rates respectively. In the range of stress around the stress plateau determined in controlled stress, the shear rate jumps from one branch to another with eventually long transient periods.

**SM6. Nonmonotonic stress in associative SPGs**

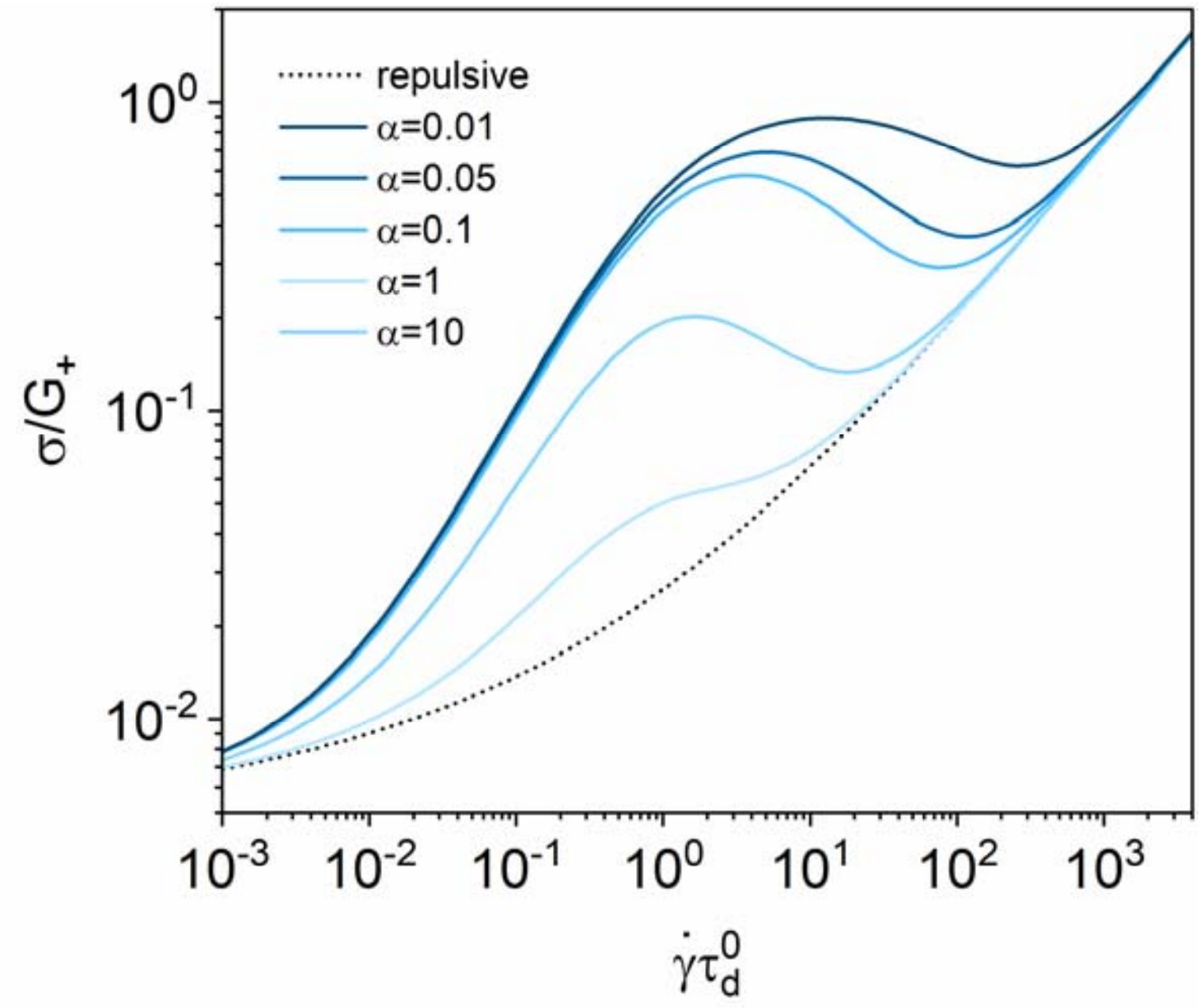

**FIG. S6** Schematic variations of the reduced stress ($\sigma = \sigma_a + \sigma_e$) versus the reduced shear rate (full line) for increasing values of α. $\sigma_a$ is computed from Eq. 12 in the main text; we take for $\sigma_e$ the stress $\sigma_e$ of the repulsive SPG at $C$ = 15 mg.g$^{-1}$ (dashed line): $\sigma_e = 1.6 + 3.5\dot{\gamma}^{0.24} + 2.1\dot{\gamma}^{0.61}$ with $G_+$ = 360 Pa.